\documentstyle[prl,aps]{revtex}

\begin{document}

\preprint{Preprint} \draft{}
\twocolumn[\hsize\textwidth\columnwidth\hsize
           \csname @twocolumnfalse\endcsname

\title{Self-organized criticality induced by quenched disorder: experiments on flux
avalanches in NbH$_x$ films.}

\author{M. S. Welling, C.~M.~Aegerter\cite{Adr}, and R.~J.~Wijngaarden
}
\address{Division of Physics and Astronomy, Faculty of Sciences,
Vrije Universiteit, De Boelelaan 1081, 1081HV Amsterdam, The Netherlands}

\date{\today}
\maketitle
\widetext

\begin{abstract}
We present an experimental study of the influence of quenched
disorder on the distribution of flux avalanches in type-II
superconductors. In the presence of much quenched disorder, the
avalanche sizes are power-law distributed and show finite size
scaling, as expected from self-organized criticality (SOC).
Furthermore, the shape of the avalanches is observed to be
fractal. In the absence of quenched disorder, a preferred size of
avalanches is observed and avalanches are smooth. These
observations indicate that a certain minimum amount of disorder is
necessary for SOC behavior. We relate these findings to the
appearance or non-appearance of SOC in other experimental systems,
particularly piles of sand.
\end{abstract}
%\pacs{DRAFT VERSION: NOT FOR DISTRIBUTION}

%\begin{multicols}{2}
\narrowtext ]

Self-organized criticality (SOC) \cite{SOC} has generated great
interest over the last 15 years mainly due to its wide range of
possible applications in many non-equilibrium systems
\cite{bak,jensen}. However, progress has been hampered by the fact
that clear, tell-tale signs of criticality, such as finite-size
scaling in the distribution of avalanches, have only been observed
in very few controlled {\em experiments} \cite{jaeger}. Recently
however, there have been a number of experimental observations of
criticality in both two- \cite{frette,ernesto} and three
dimensional systems \cite{rice,beads,socsup}. However, the
critical ingredients to obtain SOC in an experimental system still
remain obscure.

Recent theoretical advancements have studied the nature of the
criticality in SOC and made a link with phase transitions
describing how a moving object comes to rest
\cite{dickman,paczuski}. Such absorbing state phase transitions
are closely related to the roughening of an elastic membrane in a
random medium \cite{alava}. In these theoretical models, there
needs to be an absorbing state phase transition underlying the
process at hand in order to observe SOC. This critical state is
reached by a self-organization process \cite{zapperi}, which
depends on slowly driving the system away from its state where
everything is at rest. If the driving is not slow enough, the
relaxations to the critical point may be disturbed by the driving,
such that no critical state is reached \cite{grinstein}. Too
strong driving may have occurred in some of the early experiments,
particularly those performed in rotating drums and where the
grains were dropped from a considerable height. The absence of an
underlying phase transition, however, would be more fundamental
and detrimental to SOC. In the case of absorbing state phase
transitions, it is known that the presence of quenched disorder
plays an important role in the nature of the critical point, such
that this may also be an important ingredient in obtaining SOC
behavior.

%From experiments on two dimensional piles of rice with grains of
%different aspect ratios \cite{frette}, it is thought that the
%suppression of kinetic effects in the grain dynamics is important
%to obtain SOC behaviour. However, it has been shown later on that
%also round grains can lead to SOC \cite{ernesto}, if the base of
%the pile is prepared to be random, thus introducing quenched
%disorder into the system. In addition, we note that also the
%'round' grains of ref.~\cite{frette} showed finite size scaling
%and that a power-law and a stretched exponential are very
%difficult to distinguish in data covering a little more than one
%order of magnitude. Moreover, we will show below that also in
%systems where the 'grains' are overdamped, SOC behaviour may be
%absent.

Here we present an {\em experimental} investigation of the
influence of disorder on the appearance or non-appearance of SOC.
Therefore it is necessary to have a system, where the quenched
disorder can be experimentally changed while leaving all other
aspects the same, as well as having a system which has been shown
to show SOC at least in some circumstances. We study the avalanche
dynamics of magnetic vortices in the type-II superconductor
Niobium in the presence of Hydrogen impurities using
magneto-optics. As first noted by de Gennes \cite{degen}, the
penetration of a slowly ramped magnetic field into a type-II
superconductor has a strong analogy to the growing of a sand pile,
the archetypal example of SOC.
%In this analogy, the magnetic vortices take the part of the
%grains in the sand pile, the friction between grains is modelled
%by the pinning of vortices by impurities in the superconductor and
%the role of gravitation is played by the Lorentz forces between
%vortices.
It has been shown in the past, that these vortex avalanches are
distributed according to a power-law \cite{field} and more
recently that they obey finite-size scaling \cite{socsup}.
However, due to the presence of pinning in the system, studying
magnetic vortices allows a detailed investigation of the influence
of quenched disorder on their avalanche distribution and
structure. Nb is particularly suited for this purpose, as it
easily takes up H impurities (which locally destroy
superconductivity, see below)\cite{nbh}, which can be added to the
sample via a contact gas, thus allowing an experimental control of
the amount of quenched disorder in the system. Furthermore,
studying magnetic vortices has the advantage that kinetic effects
in their dynamics, which are thought to have hampered some
sand-pile experiments \cite{frette}, are naturally absent due to
the over-damped dynamics of the vortices \cite{blatterbible}.

The experiments described here were carried out on a Nb film of a
thickness of 500 nm, evaporated under ultra high vacuum conditions
on an 'R-plane' sapphire substrate. The films were then covered
with a 10 nm Pd caplayer in order to facilitate the catalytic
uptake of H into the film \cite{marco}.

The local magnetic flux density, $B_z$, just above the sample is
measured using an advanced magneto-optical (MO) setup, directly
yielding the local Faraday-angle in an Yttrium-Iron Garnet
indicator, using a lock-in technique \cite{rsi}. Images are taken
with a charge-coupled device camera (782 $\times$ 582 pixels) with
a resolution of 3.4 $\mu$m per pixel. The sample is placed in a
cryo-magnet and cooled to 4.2 K in zero applied field. The
external field is then ramped from 0 to 20 mT in steps of 50
$\mu$T, where the flux in the sample is relaxed for 3 seconds
before an MO picture is acquired. This sequence is repeated twice
for each H concentration to check for reproducibility.

The quenched disorder in the sample is increased after such a
sequence of two field sweeps by equilibrating the sample with a
certain higher partial pressure of H at room temperature for one
hour. The partial loading pressures used in the experiments
discussed here were 80, 260, 1130, and 1810 Pa. We estimate that
the H impurities present in the as grown sample correspond to a
partial loading pressure lower than roughly 10 Pa. After
equilibration, ensuring a uniform distribution of H in the Nb
film, the sample is cooled down again. In this cooling procedure,
phase separation occurs in H-rich and H-poor regions
\cite{epsilon}, where superconductivity is suppressed in the
H-rich phase \cite{vinnikov}. Thus the clusters of H-rich phase,
which are in the order of 0.1 to 1 $\mu$m in diameter
\cite{epsilon}, act as effective pinning sites for the vortices
and thereby introduce quenched disorder.

\begin{figure}
\input{epsf}
\epsfxsize 8cm %\epsfysize 6.5cm
\centerline{\epsfbox{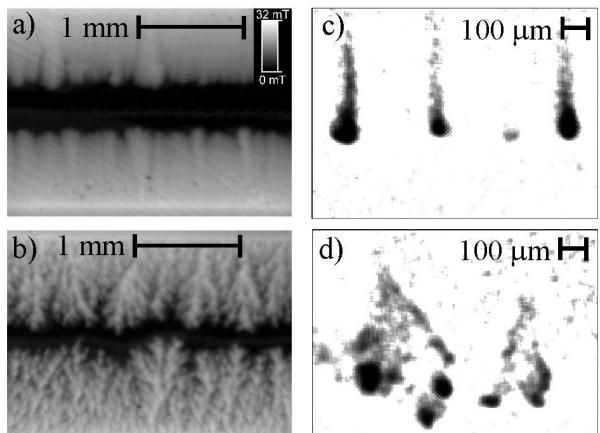}} \caption[~]{a) The magnetic flux
landscape in the center of the Nb film without added Hydrogen
impurities. The magnetic flux density in the sample is indicated
by the brightness, where white corresponds to 32 mT and black
corresponds to 0 mT (see inset). b) The flux landscape in the same
sample after the application of a H pressure of 1810 Pa at room
temperature. As can be seen, the flux structures are much more
disordered and branched. c) Magnetic flux avalanches in the
absence of H treatment, as given by the subtraction of two
consecutive MO images. Note that the magnification and the
grey-scale is different from a), white corresponds to 0 mT and
black corresponds to 3 mT. The avalanches have a predominant size
and are rather smooth and plume-shaped. d) Flux avalanches after H
treatment with 1810 Pa gas pressure at room temperature, again
obtained from a subtraction of consecutive MO images (grey scale
and magnification are as in c). The avalanches are more fractal
and branched than in c) and do no longer have a preferred size
(see text).} \label{struct}
\end{figure}

In this manner, we obtain a collection of magnetic flux landscapes
as shown in Fig.~\ref{struct}a), which are then analyzed in terms
of the avalanches that have occurred between time steps. In order
to obtain the amount of magnetic flux that has been transported in
one avalanche, we first average each image over a set of 4
$\times$ 4 pixels to reduce noise. Then two consecutive images are
subtracted leading to an image like that shown in
Fig.~\ref{struct}c). As can be seen in the figure, in each image
we observe a number of avalanches. These are identified
individually, using a threshold of 0.3 mT in $\Delta B_z$. The
amount of flux in each of these avalanches is subsequently
determined from integrating the difference in flux density over
the area of the avalanche:
\begin{equation}
s = \Delta \Phi = \int \Delta B_z dxdy.
\end{equation}
From the spatial resolution of the setup and the threshold level
of the avalanche identification, we can determine the smallest
avalanches that are still resolved to contain about 20 $\Phi_0$,
where $\Phi_0 = h / 2e$ is the magnetic flux carried by each
vortex. In order to check for finite size scaling, we discard
avalanches exceeding a linear extent of 200, 100, and 50 $\mu$m
from the analysis. This corresponds to decreasing the system size
accordingly, as the linear extent of an avalanche cannot exceed
the system size.

\begin{figure}
\input{epsf}
\epsfxsize 8cm %\epsfysize 6.5cm
\centerline{\epsfbox{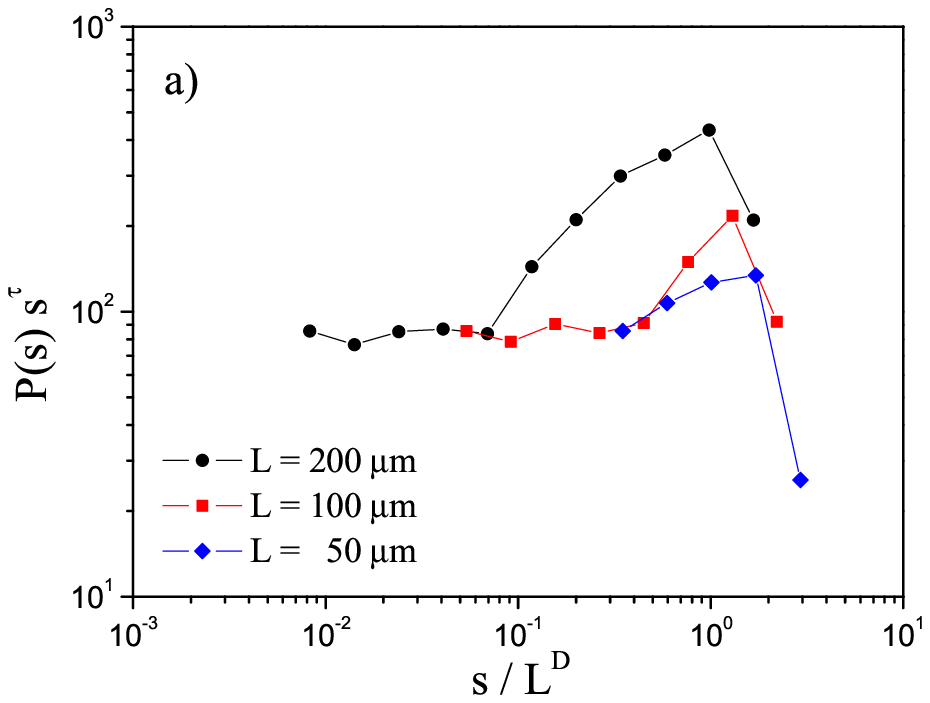}} \epsfxsize 8cm %\epsfysize 6.5cm
\centerline{\epsfbox{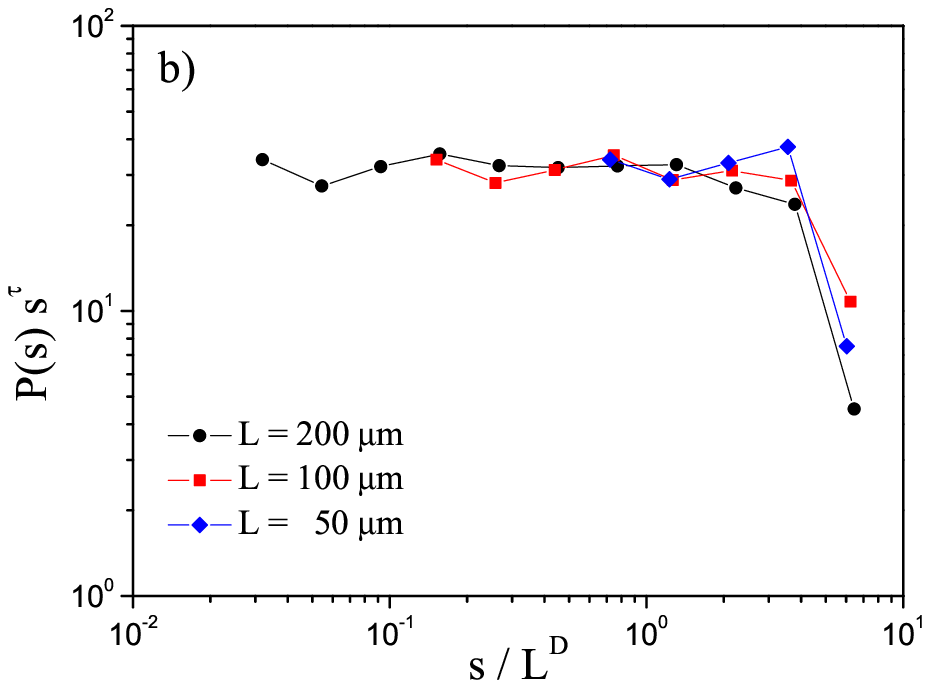}} \caption[~]{a) The flux
avalanche size distributions in the absence of Hydrogen treatment.
Data are shown for sets of avalanches not exceeding different
linear sizes of $L$ = 200, 100, and 50 $\mu$m. The different
distributions are vertically scaled with $s^\tau$ and the sizes
are horizontally scaled with $L^{-D}$, in order to obtain the
usual curve collapse for finite size scaling. However, as
avalanches of $\sim$2500 $\Phi_0$ are preferred, no good curve
collapse is obtained in this region. For small avalanches and
avalanches of the cut-off size, the curves do collapse, such that
the fractal dimension of the avalanches can be determined. b) The
avalanche size distributions scaled in the same way, for
experiments after H treatment at 1810 Pa of the sample. Here a
good curve collapse is observed, with power-law scaling over more
than two decades. This is clear evidence that in the presence of
quenched disorder, SOC is present in the vortex avalanches.}
\label{fss}
\end{figure}

As can be seen in Fig.~\ref{struct}c), when the sample has not
been treated with H, the avalanches have a characteristic
structure and area. This leads to a peak in the size distribution,
which disappears as avalanches exceeding the preferred linear
extent are discarded. Thus in the absence of H-induced quenched
disorder, no SOC behavior is observed, as there is no finite-size
scaling of the avalanche distribution. This is shown in
Fig.~\ref{fss}a), where we show the scaled avalanche size
distribution for the sample without H treatment. Here, the size
distribution has been logarithmically binned, such that avalanche
size steps in the histogram are separated by a constant factor
rather than a constant step width. Moreover, the histograms are
scaled with $s^\tau$, which leads to horizontal lines in case the
avalanches are power-law distributed. The displaced flux per
avalanche is scaled by $L^{-D}$, where $L$ is the scale above
which avalanches are discarded and $D$ is the fractal dimension of
the avalanches. In order to obtain the 'collapse' in
Fig.~\ref{fss}a), we have used $\tau = 1.35$ and $D = 2.75$. The
value of $\tau$ is determined from a best collapse of the data
onto a horizontal line for small s. The value of $D$ is found from
the best collapse of the cut-off values. The nice collapse of
cut-off values indicates that the avalanches in this case do have
a definite fractal dimension, which is close to 3, as one would
expect given their smooth appearance in Fig.~\ref{struct}c). The
peak that can be observed in the distribution is due to the
predominance of avalanches containing about $\sim$ 2500 $\Phi_0$,
which can also be inferred from Fig.~\ref{struct}c).

Turning to the data in Fig.~\ref{struct}d), it can be seen that in
the presence of strong quenched disorder (i.e. after treating the
sample with 1810 Pa of H gas), the avalanche structure is markedly
different from that in the pristine case. Now, the flux moves in a
much more branched way and furthermore a preferred size of
avalanches is absent. Again, we have checked this with a
finite-size scaling analysis of the avalanche size distributions,
see Fig.~\ref{fss}b). The size distribution is again
logarithmically binned and scaled in order to obtain a curve
collapse. Here, in contrast to Fig.~\ref{fss}a), there is good
curve collapse and thus we can determine the size distribution
exponent and avalanche dimension to be $\tau = 1.07(2)$ and $D =
2.25(5)$, where we observe a power-law distribution over more than
two orders of magnitude.

This overall behavior of flux avalanches in the presence of
changeable quenched disorder has also been studied in molecular
dynamics simulations of a collection of vortices \cite{olson}.
This is a great advantage of the vortex system over the study of
sand piles, as the interactions between vortices are well known
\cite{blatterbible} and can thus be simulated in order to separate
collective effects from microscopic dynamics. In their study,
Olson {\em et al.} \cite{olson} have found that on increasing the
pinning density, the avalanche distributions change from showing a
preferred size (with a power-law distribution at small sizes) to a
power-law distribution. Furthermore, in the case of strong
pinning, the exponent $\tau$ obtained in these power-law
distributions is between 1 and 1.4 in very good agreement with our
experimental values. Moreover, in molecular dynamics simulations,
the movements of all vortices can be followed and it turns out
that in the case of low pinning density, the avalanches form
channels \cite{olson}, which is highly reminiscent of the
structures seen in Fig.~\ref{struct}c).

%In a coarse-grained model of flux penetration into superconductors
%\cite{basslerpac}, the changing of the pinning density was also
%studied. However in that case, no effect was found. A possible
%reason for this is that the model is applicable for vortex
%densities which are much higher than those in our experimental
%situation, thus affecting the effective pinning density.

Let us now turn to the general implications on the appearance of
SOC that can be drawn from the above observations. As we have
shown, in the absence of quenched disorder, even the non-kinetic
vortices \cite{blatterbible} do not show finite-size scaling in
their avalanche distribution. Furthermore, it has been shown
previously that steel balls in a two dimensional pile can show SOC
\cite{ernesto} when quenched noise is introduced via a fixed
random arrangement of balls at the base. These two observations
together seem to indicate that kinetic effects may be less
important than previously thought \cite{frette}, while it is the
presence of quenched noise that leads to the appearance of SOC.
This is also corroborated by molecular dynamics simulations on
vortices, using over-damped dynamics finding the same result
\cite{olson}.

%We would also like to note that the 'round' grains of
%ref.~\cite{frette} did show finite size scaling and were only
%considered not to show SOC because the form of the distribution
%was that of a stretched exponential. However, a power-law and a
%stretched exponential are very difficult to distinguish in data
%covering a little more than one decade, such this may well have
%constituted an observation of SOC as well. As was the case in
%\cite{ernesto}, this could be due to the introduction of quenched
%disorder by fixing a random distribution of rice to the bottom
%plate.

\begin{figure}
\input{epsf}
\epsfxsize 8cm %\epsfysize 6.5cm
\centerline{\epsfbox{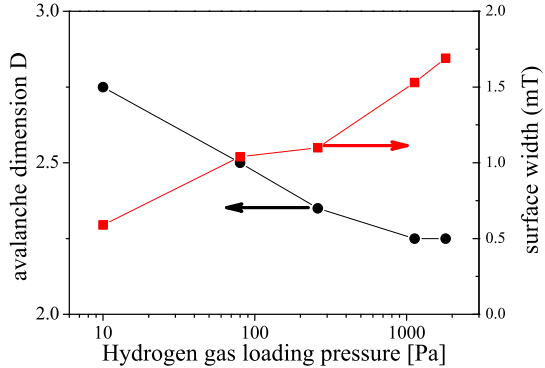}} \caption[~]{The fractal dimension
of the avalanches (circles) and the surface width (see text) of
the flux landscape (squares) as a function of Hydrogen partial
pressure. Data for the pristine films are shown at 10 Pa, the
estimated maximum contamination H pressure. The fractal dimension
decreases with increasing quenched disorder, corresponding to a
more branched and rough structure, as seen in Fig.~\ref{struct}d).
The surface width increases with quenched disorder, which
indicates that the flux landscape becomes increasingly rough as H
is introduced into the sample, as can be seen in
Fig.~\ref{struct}b). } \label{rough}
\end{figure}

The importance of quenched disorder can be best seen in the
structure of the avalanches, which is intimately connected to the
surface of a pile or flux landscape \cite{rice,pacz}. As the
number of pins is increased, the vortex avalanches have to
accommodate to their presence. This frustration leads to a much
more branched and open avalanche structure. This is shown in
Fig.~\ref{rough}, where the circles show the values of $D$ for
different H loading pressures. As can be seen, at low pinning
density, the avalanches have a dimension close to 3, indicating a
smooth structure, whereas with increasing H content, the avalanche
dimension decreases. Due to the more complex structure of an
avalanche, a bigger range of possible avalanche sizes can be
explored over the area of the sample. Thus there is no longer a
preferred size of avalanches and hence SOC behavior occurs. Again,
we can compare our results to those of Altshuler {\em et al.}
\cite{ernesto}, where in the absence of SOC, only a small part of
the pile took part in avalanches and the pile surface overall was
much smoother than in the case where SOC was observed.

In order to quantify the roughness \cite{barabasi} of the flux
landscape, we determine the width, $\langle (B_z(x,y) - \langle
B_z(x,y) \rangle_y)^2 \rangle^{1/2}$, of the flux landscape after
subtraction of the average profile, $\langle B_z(x,y) \rangle_y$.
As can be seen in Fig.~\ref{rough}, with increasing H content, the
flux landscape becomes more rough as given by a fourfold increase
in surface width. Thus as has been observed in experiments using
steel balls \cite{ernesto} as well as in rice piles
\cite{rice,oslo}, SOC behavior in the flux landscape is coupled to
the observation of a rough pile surface.

%This connection between the avalanche structure and the surface
%roughness is also indicated by the fact that the roughness
%exponent, $\alpha$, describing the scaling of the interface width
%in space, is closely connected to the avalanche properties. In the
%case of high pinning density, the fractal dimension of the active
%area of the avalanches is determined to be $d_B = 1.5(1)$, which
%leads to a roughness exponent of $\alpha = D - d_B$ = 0.75(10).
%This is in good agreement with a determination via the correlation
%function as described above, where we obtain $\alpha$ = 0.85(5).

In conclusion, we have shown in the case of high pinning density
that the magnetic flux avalanches observed in a Nb film show
finite-size scaling, which implies the presence of SOC in the
system. However, in the presence of a low pinning density, a
preferred size of avalanches is found and no finite-size scaling
is observed. This demonstrates the importance of quenched disorder
in the system in order to obtain SOC behavior. The exact position
of the transition from non-SOC to SOC behavior cannot be
determined very accurately due to experimental limitations (e.g.
'noise' in Fig. 2 and limited accessible range of $L$), but might
also be intrinsically smooth if one considers the gradual changes
observed in Fig. 3 for the avalanche dimension $D$ and surface
width. Our {\em experimental} observation of this transition as a
function of disorder is in agreement with previous numerical work
on cellular automata \cite{puhl} and with molecular dynamics
simulations of magnetic vortices \cite{olson}. In addition, our
findings are consistent with the point of view of SOC as an
absorbing state phase transition \cite{dickman,zapperi}. In
absorbing state phase transitions, such as directed percolation
\cite{barabasi}, the density of quenched disorder is a critical
parameter, whereby a phase transition can be induced. The presence
of such an underlying phase transition is a necessary condition in
order to obtain SOC in the model of Vespigniani {\em et al.}
\cite{zapperi}. Thus again, this view advocates that the increase
of quenched noise can lead to the appearance of SOC, as indeed we
have found experimentally.

%Furthermore, the connection between roughening physics and SOC is
%put on a firm experimental basis, as the appearance of SOC goes
%together with increased roughening of the flux landscape.

We would like to thank Ruud Westerwaal for help with the
preparation of the samples. This work was supported by FOM
(Stichting voor Fundamenteel Onderzoek der Materie), which is
financially supported by NWO (Nederlandse Organisatie voor
Wetenschappelijk Onderzoek).

\bibliographystyle{prsty}

\begin{thebibliography}{10}

\bibitem[*]{Adr} Present address: Fachbereich Physik, Universit\"at Konstanz,
P.O. Box 5560, 78457 Konstanz, Germany.
\bibitem{SOC} P.~Bak, C. Tang, and K. Wiesenfeld, Phys. Rev. Lett.
{\bf 59}, 381 (1987) and Phys. Rev. A {\bf 38}, 364 (1988).
\bibitem{bak} P.~Bak, {\em How nature works} (Oxford Univ. Press, 1995).
\bibitem{jensen} H.-J. Jensen, {\em Self-Organized Criticality}
(Cambridge University Press, 2000).
\bibitem{jaeger} see e.g. H.~M.~Jaeger, R. P. Behringer, and S. R. Nagel, Rev. Mod. Phys.
{\bf 68}, 1259 (1996); S.~R.~Nagel, {\em ibid.} {\bf 64}, 321
(1992) and references therein.
\bibitem{frette} V.~Frette {\em et al.}, Nature (London) {\bf 379}, 49 (1996).
\bibitem{ernesto} E. Altshuler {\em et al.}, Phys. Rev. Lett.
{\bf 86}, 5490 (2001).
\bibitem{rice} C. M. Aegerter, R. G\"unther, and R. J. Wijngaarden, Phys. Rev. E {\bf 67}, 051306 (2003).
\bibitem{beads} R. M. Costello {\em et al.}, Phys. Rev. E {\bf
67}, 041304 (2003).
\bibitem{socsup} C. M. Aegerter, M. S. Welling, and R. J. Wijngaarden, Europhys.
Lett. {\bf 65}, 573 (2004).% cond-mat 0305591.
\bibitem{dickman} R. Dickman, {\em et al.}, Braz. J. Phys. {\bf 30}, 27 (2000); cond-mat 9910454.
\bibitem{paczuski} M. Paczuski, S. Maslov, and P. Bak, Europhys.
Lett. {\bf 27}, 97 (1994).
\bibitem{alava} M. J. Alava and K. B. Lauritsen, Europhys. Lett. {\bf 53}, 563 (2001);
G. Pruessner, Phys. Rev. E {\bf 67}, 030301 (2003).
\bibitem{zapperi} A. Vespignani and S. Zapperi, Phys. Rev. Lett.
{\bf 78}, 4793 (1997); Phys. Rev. E {\bf 57} 6345 (1998); R.
Dickman {\em et al.}, Phys. Rev. E {\bf 57}, 5095 (1998).
\bibitem{grinstein} G. Grinstein, J. Appl. Phys. {\bf 69}, 5441
(1991).
\bibitem{degen} P. G. de Gennes, {\em Superconductivity of metals
and alloys} (Addison-Wesley, New York, 1966).
\bibitem{field} S. Field, J. Witt, and F. Nori, Phys. Rev. Lett. {\bf 74},
1206 (1995); C. M. Aegerter, Phys. Rev. E {\bf 58}, 1438 (1998);
K. Behnia {\em et al.}, Phys. Rev. B {\bf 61}, R3815 (2000); E.
Altshuler {\em et al.}, cond-mat/0208266.
\bibitem{nbh} G. Alefeld and J. V\"olkl, {\em Hydrogen in Metals
II, Topics in applied Physics vol. 29} (Springer, Heidelberg,
1978).
\bibitem{blatterbible} G. Blatter {\em et al.}, Rev. Mod. Phys.
{\bf 66}, 1125 (1995).
\bibitem{marco} M. S. Welling {\em et al.}, Physica C {\bf 406}, 100
(2004).
\bibitem{rsi} R. J. Wijngaarden {\em et al.} Rev. Sci. Instrum.
{\bf 72}, 2661 (2001).
\bibitem{epsilon} T. Schober, Phys. Stat. Solidi {\bf 30}, 107
(1975); K. N\"orthemann {\em et al.}, J. Alloys Comp. {\bf
356-357}, 541 (2003).
\bibitem{vinnikov} E. G. Maksimov and O. A. Pankratov, Usp. Fiz. Nauk
{\bf 116}, 403 (1975); L. Ya. Vinnikov {\em et al.}, Sov. J. Low
Temp. Phys. {\bf 3}, 4 (1977).
\bibitem{olson} C. J. Olson, C. Reichhardt, and F. Nori, Phys. Rev. B {\bf 56},
6175 (1997).
%\bibitem{basslerpac} K. E. Bassler and M. Paczuski, Phys. Rev.
%Lett. {\bf 81}, 3761 (1998).
\bibitem{pacz} M. Paczuski, S. Maslov, and P. Bak, Phys. Rev. E
{\bf 53}, 414 (1996).
\bibitem{barabasi} A. L. Barabasi and H. E. Stanley {\em Fractal Concepts in
Surface Growth} (Cambridge University Press, 1995).
\bibitem{oslo} A. Malthe-S$\o$renssen {\em et al.}, Phys. Rev. Lett. {\bf 83}, 764 (1999).
\bibitem{puhl} H. Puhl, Physica A {\bf 197}, 14 (1993).
%\bibitem{river} V. B. Sapozhnikov, and E. Foufoula-Georgiou, Water
%Res. Res. {\bf 33}, 1983 (1997), {\em ibid} {\bf 32} 1429 (1996),
%and {\em ibid} {\bf 32} 1109 (1996).
%\bibitem{boettcher} M. Paczuski, and S. Boettcher, Phys. Rev.
%Lett. {\bf 76}, 348 (1996).
%\bibitem{alava2} M. J. Alava, J. Phys. Cond. Mat. {\bf 14(9)}, 2353 (2002).

\end{thebibliography}

%\end{multicols}
\end{document}